\documentclass[pre,showpacs,twocolumn,floatfix,amsmath,amsfonts]{revtex4}
\usepackage{graphicx,subfigure}
\usepackage{epstopdf}
\usepackage{textcomp}
\usepackage[utf8]{inputenc}
\usepackage[russian,english]{babel}
\usepackage[usenames, dvipsnames]{color}
\usepackage{hyperref}
\usepackage{color}

\begin{document}

\title{Mass transfer in Frenkel-Kontorova chain initiated by molecule impact}

\author{A.~Moradi~Marjaneh$^1$}
\email{moradimarjaneh@gmail.com}
\author{D.~Saadatmand$^2$}
\author{I.~Evazzade$^3$}
\author{R.~I.~Babicheva$^4$}
\author{E.~G.~Soboleva$^5$}
\author{N. Srikanth$^6$}
\author{Kun~Zhou$^2$}
\author{E.~A.~Korznikova$^{7,8}$}
\author{S.~V.~Dmitriev$^{7,9}$}
\vspace{6mm}

\affiliation{$^1$Young Researchers and Elite Club, Quchan Branch, Islamic Azad University, Quchan, Iran
\\
$^2$Department of Physics, University of Sistan and Baluchestan, Zahedan, Iran \\
$^3$Department of Physics, Ferdowsi University of Mashhad, Mashhad, Iran\\
$^4$School of Mechanical and Aerospace Engineering, Nanyang Technological University, 639798, Singapore \\
$^5$Yurga Institute of Technology (Branch), National Research Tomsk Polytechnic University, 652050 Yurga, Russia\\
$^6$Interdisciplinary Graduate School, Nanyang Technological University, Singapore 639798, Singapore\\
$^7$Institute for Metals Superplasticity Problems, Russian Academy of Sciences, Ufa, 450001 Russia \\
$^8$Ufa State Aviation Technical University, 450008, Ufa, Russia \\
$^9$National Research Tomsk State University, Lenin Avenue 36, 634050 Tomsk, Russia
}

\date{\today}

\begin{abstract}
The Frenkel-Kontorova chain with a free end is used to study initiation and propagation
of crowdions (anti-kinks) caused by impact of a molecule consisting of $K$ atoms.
It is found that molecules with $1<K<10$ are more efficient in initiation of crowdions
as compared to single atom ($K=1$) because total energy needed to initiate the crowdions
by molecules is smaller. This happens because single atom can initiate in the chain
only sharp, fast-moving crowdions that requires a relatively large energy. Molecule has
finite length, that is why it is able to excite a wider crowdion with a smaller velocity
and smaller energy. Our results can shed light on the atomistic mechanisms of mass transfer
in crystals subject to atom and molecule bombardment.
\end{abstract}


\maketitle

\section{Introduction}

Bombardment of crystal surface by ionized or neutral atoms or molecules is a phenomenon observed
either at ambient conditions or during technological surface treatment such as ion implantation,
plasma surface treatment, magnetron sputtering, etc. \cite{n1,n2,n3,n4,n5,n6,n7}. As a result, desired or undesired structure
transformations near crystal surface can take place due to the mass transfer inside the crystal
initiated by the bombardment.

Point defects such as vacancies and interstitial atoms play a very important role in the physics of crystalline
solids transporting mass during plastic deformation \cite{1,2,3,4,5,6,7}, irradiation
\cite{9,10,11,12,13}, heat treatment \cite{8,n8}, etc. Thermally activated diffusion mainly occurs
through vacancy migration mechanism \cite{8,n8}. Energy of interstitial atoms is larger, therefore
their concentration in thermal equilibrium is much smaller than that of vacancies. The role of
interstitials largely increases in far-from-equilibrium processes with energy flux through the crystal.
Interstitials can be immobile \cite{14} or mobile, in the latter case they are located in close-packed
atomic rows in the form of crowdions \cite{15}. Very often crowdions have lower potential energy than
immobile interstitials \cite{15,16}. Crowdions can be at rest or they can move with a speed below or
above the speed of longitudinal sound \cite{17,18,19}. Standing or subsonic crowdions have a kink
profile in a close-packed atomic row, spanning over half a dozen of atoms. However, supersonic crowdions are
highly localized with only one or two atoms moving with a high speed at the same time \cite{19,20}.

Crowdions can be viewed as topological solitons being very efficient in mass and energy transport
\cite{6,21,22,23,24,25,26,27,28}. Moving excitations in crystals are actively studied in order to
to explain various nontrivial experimental results such as annealing of defects deep inside
germanium single crystal by surface plasma treatment \cite{29} or tracks in mica
crystals \cite{19,30,31,32,33,34,35}. Discrete breathers \cite{36,37,38,39}, crowdions \cite{19,40,41},
and quodons \cite{42} have been considered as movable excitation candidates in mica.
Collisions of supersonic crowdions in two-dimensional (2D) model crystals have been studied in \cite{43}.
The ability of supersonic crowdions and discrete breathers to carry electric charge has been
analyzed by Kosevich \cite{44}.

Static crowdions have been investigated using first principles simulations \cite{45,46,47},
while their dynamics have been analyzed with the use of molecular dynamics method \cite{19,48}.

Recently notion of supersonic $N$-crowdions has been introduced based on molecular dynamics
simulations \cite{Nc1,Nc2}. $N$-crowdions can carry more than one interstitial atoms along
a close-packed atomic row. For their excitation equal initial momentum was given to $N$
neighboring atoms in a close-packed row along the row. For 2D and 3D Morse crystals,
it has been shown that $N$-crowdions transport interstitial atoms more efficiently than
classical 1-crowdions because they travel longer distances having lower initial
energy~\cite{Nc1,Nc2}. On the other hand, there remains the question of knowing how $N$-crowdions can
be excited in reality. It is very difficult to construct a scenario when several neighboring atoms
in a close-packed atomic row simultaneously gain a relatively large momentum along the row
in the bulk of a crystal. On the contrary, this can be easily achieved when a molecule hits
the surface of the crystal.

In the present study, we consider simple 1D Frenkel-Kontorova model to demonstrate that
kicks by molecules are more efficient in initiation of mass transfer by crowdions than
the case of single atoms impact.

The outline of the paper is as follows.  In Sec.~\ref{SimulationSetup} the model and simulation details are described. Our main result is presented in Sec.~\ref{Numerics}, where bombardment of the Frenkel-Kontorova chain with single atoms and molecules is simulated. In order to better understand the results of Sec.~\ref{Numerics}, we analyse properties of anti-kinks (crowdions) in Sec.~\ref{KinkSolution}. Finally, conclusions are presented in Sec.~\ref{Conclusions}.

\section{The model and simulation setup}
\label{SimulationSetup}

The main results of this study are obtained in the frame of 1D Frenkel-Kontorova model described
in Sec.~\ref{Model}. In the first place, we analyse the 3D fcc Morse crystal in Sec.~\ref{Model3D} to justify the choice of parameters of the 1D model.

\subsection{3D Morse crystal}
\label{Model3D}

\begin{figure}
\includegraphics[width=6cm,height=3.7cm]{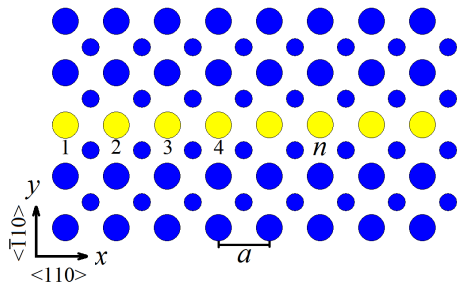}
\caption{Atoms of fcc lattice shown in $xy$-projection with the $x$, $y$, and $z$ axes along
$\langle 110 \rangle$, $\langle \overline{1}10 \rangle$, and $\langle 001 \rangle$
close-packed crystallographic directions, respectively. Atoms of two neighboring atomic planes
parallel to $xy$-plane are shown with circles of different size. $a$ is the interatomic
distance. Atoms of one close-packed row (shown by yellow circles) are numbered with index $n$.}
\label{fig1}
\end{figure}

The fcc lattice with the lattice parameter $d$ and interatomic distance $a=d/\sqrt 2$ is considered.
As shown in Fig.~\ref{fig1}, Cartesian coordinate system is used with the $x$, $y$, and $z$ axes oriented
along $\langle 110 \rangle$, $\langle \overline{1}10 \rangle$, and $\langle 001 \rangle$
close-packed crystallographic directions, respectively.

\begin{figure}
\includegraphics[width=9cm,height=7cm]{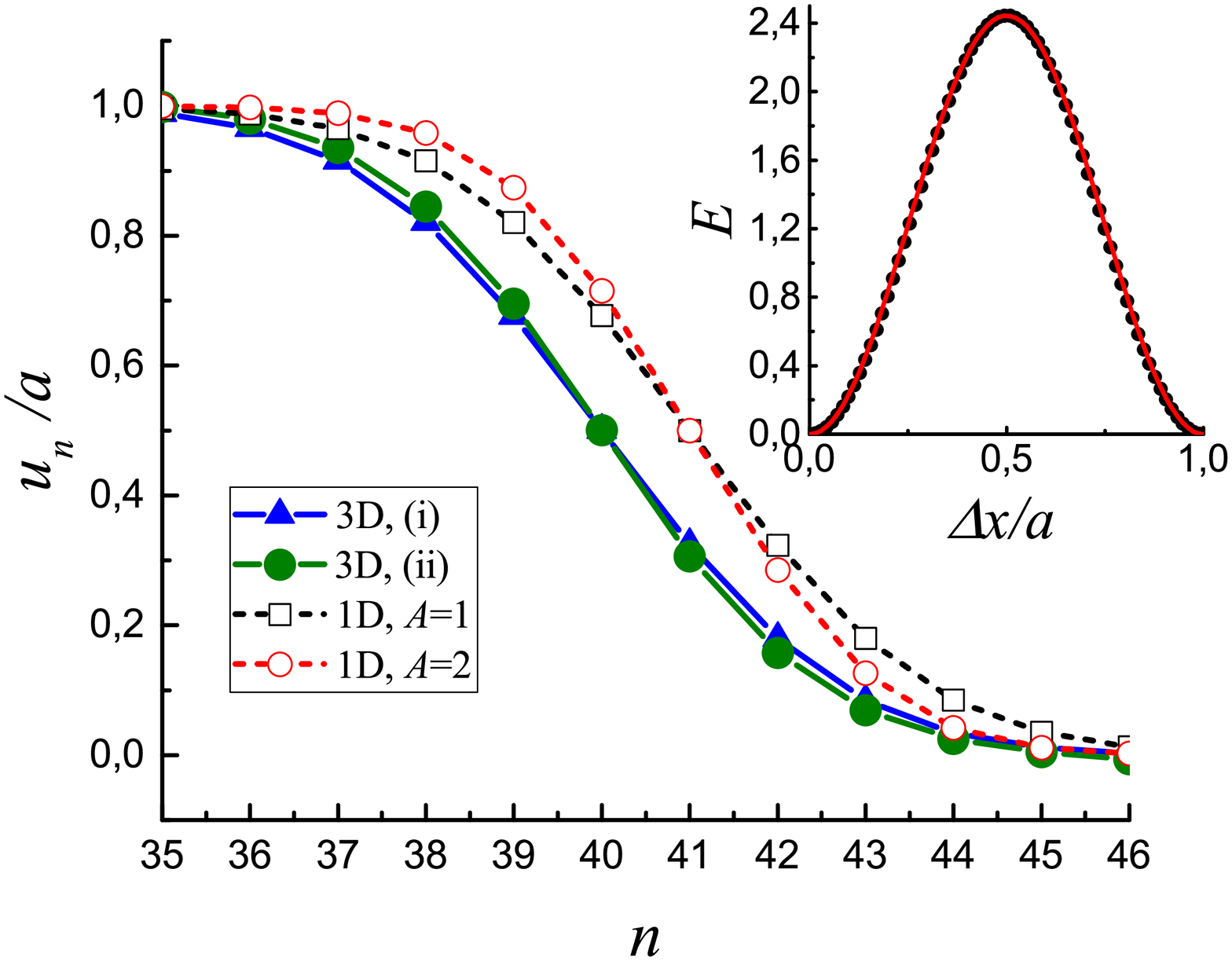}
\caption{Relaxed crowdion profiles. Triangles and filled circles show the results for 3D fcc Morse
lattice, cases (i) and (ii), respectively (see in the text). Open squares and open circles are
for the 1D Frenkel-Kontorova model, for $A=1$ and 2, respectively. Inset shows the effective
on-site potential created by the atoms of 3D fcc lattice for a close-packed atomic row.
Dots show the numerical result and the line is the fit $E=A[1-\cos(2\pi\Delta x/a)]$ with $A=1.22$.} \label{Kinks}
\end{figure}

Atoms interact via classical Morse pair potential \cite{Morse}
\begin{eqnarray}\label{Morse}
U(\xi) = D\left(1+e^{-2\alpha(\xi-r_m)}-2e^{-\alpha(\xi-r_m)}\right),
\end{eqnarray}
where $U$ is the potential energy of two atoms at a distance $\xi$ apart,
$D$ is the depth of the potential (bond energy), $U$ has a minimum at the equilibrium distance $\xi=r_m$,
and $\alpha$ defines the bond stiffness. We use dimensionless units and, without loss of generality,
set atom mass $m$ equal to 1 and
\begin{eqnarray}\label{Dm}
D=1, \quad r_m=1.
\end{eqnarray}
For the bond stiffness we set the typical value of
\begin{eqnarray}\label{alpha}
\alpha=4.
\end{eqnarray}
With this choice, $U(\xi)$ is negligibly small for $\xi>5r_m$ so that this value is taken as the cut-off radius.
The equilibrium interatomic distance in this case is $a$=0.90142.


Thermal fluctuations are not taken into account, i.e., simulations are done at 0~K.

The computational cell contains 3840 atoms having dimensions $40a\times 8a\times 12a/\sqrt 2$.
Periodic boundary conditions are used.

Atoms of one close-packed atomic row parallel to the $x$ axis are numbered with index $n$
as shown in Fig.~\ref{fig1}. In this row, in the center of the computational cell,
we create a crowdion (anti-kink) using the following ansatz
\begin{eqnarray}\label{chainposition0}
u_{n}=\frac{a}{2}\{1-\tanh[\beta(n-x_0)]\},
\end{eqnarray}
where $u_{n}$ is the initial displacement of $n$-th atom along the $x$ axis, $\beta=0.3$ and $x_0=40$
are the crowdion inverse width and initial position, respectively. Initial velocities of all atoms
in the computational cell are equal to zero. Note that application of the ansatz
(\ref{chainposition0}) makes the cite $n=0$ vacant.

We then apply the relaxation procedure to find the equilibrium configuration of the on-site crowdion.
This procedure is done for two cases: (i) all atoms in the computational cell are movable and (ii) only atoms in the close-packed
atomic row containing the crowdion are movable. In the latter case, the atoms surrounding the row with crowdion
create a rigid on-site potential, while in the former case they are free to relax.

In Fig.~\ref{Kinks}, we present the relaxed crowdion profiles for the case (i) by triangles and for the case (ii) by filled circles.
As expected, crowdion width is smaller in the case (ii).

Our next step is to calculate the on-site potential created by the atoms surrounding a close-packed
atomic row. To do so, we shift the close-packed atomic row as a rigid body along $x$ axis by $\Delta x$
and calculate the potential energy of an atom of the row, $E$. In the inset of Fig.~\ref{Kinks}, we plot $E$
as the function of $\Delta x/a$ by dots. The height of the potential is $2.44\equiv 2A$. By a solid line,
we show the sinusoidal function $E=A[1-\cos(2\pi\Delta x/a)]$ of amplitude $A=1.22$. Maximal difference between the numerical data and
the sinusoidal fit is less than 2.5\%. This estimation of the on-site potential height will be used in the formulation
of the 1D Frenkel-Kontorova model.

\begin{figure}
\includegraphics*[width=0.5\textwidth]{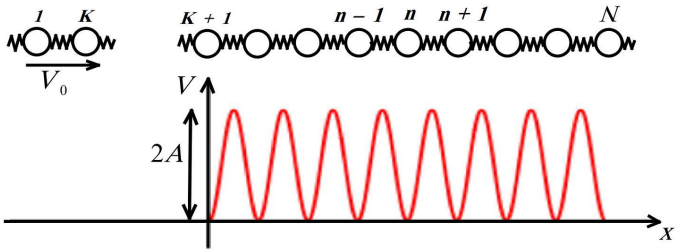}
\caption{Scheme of the simulation setup. Molecule of $K$ particles moves with the initial velocity $V_0$ and hits the end of the Frenkel-Kontorova chain of $M$ particles placed in sinusoidal on-site potential of depth $2A$. Each particle interacts with the nearest neighbors via the Morse potential. Total number of particles in the system is $N=K+M$, they are numbered by index $n$.}
\label{fig:1}
\end{figure}

\subsection{1D model}
\label{Model}

We consider the Frenkel-Kontorova chain of $M$ particles placed in the sinusoidal on-site potential. A molecule consisting of $K$ particles moves with the velocity $V_0$ and hits the left end of the chain (see Fig.~\ref{fig:1}). Total number of particles in the system is $N=K+M$. Particles interact with the nearest neighbors via Morse potential. Hamiltonian of the system is given by
\begin{eqnarray}\label{FK}
H = \sum_{n=1}^{N} \frac{m}{2}\Big(\frac{du_n}{dt}\Big)^2+\sum_{n=1}^{N-1} U(u_{n+1}-u_n) +\sum_{n=1}^{N} V(u_n),
\end{eqnarray}
where longitudinal coordinates of the particles as a functions of time, $u_n(t)$, are to be determined.

The first term in Eq.~(\ref{FK}) gives the kinetic energy of the system. We take $m=1$ for particle mass, which can always be achieved by proper choice of time unit.

The second term in Eq.~(\ref{FK}) gives the Morse interaction between nearest neighbors described by Eq.~(\ref{Morse})
with the parameters used for the 3D fcc crystal.

The on-site potential has been represented by the third term in Eq.~(\ref{FK}) which is taken in the form
\begin{equation}
 V(\eta) =
  \begin{cases}\label{Onsite}
    0       & \quad \text{for  } \eta\le0, \\
    A[1-\cos(2\pi\eta)]  & \quad \text{for  }  \eta>0.
  \end{cases}
\end{equation}
The on-site potential has amplitude $A$, its period is equal to unity to be commensurate with the inter-particle distance, and it acts only in the region $x>0$, as schematically shown in Fig.~\ref{fig:1}.

As it was shown in Sec.~\ref{Model3D}, in the 3D fcc Morse crystal $A=1.22$ (see also inset in Fig.~\ref{Kinks}).
In order to see the effect of $A$, we consider two values for the amplitude of the on-site potential in the 1D model, namely, $A=1$ and $A=2$.
Static anti-kink profiles calculated numerically for the 1D model are shown in Fig.~\ref{Kinks} by open squares for $A=1$ and by open
circles for $A=2$. It can be seen that they have slopes close to the slopes
of crowdions in 3D fcc crystal. Naturally the anti-kink slope is larger for larger $A$.


From the Hamiltonian specified by Eqs.~\eqref{FK}, (\ref{Morse}), and \eqref{Onsite}, the following equations of motion can be
derived
\begin{eqnarray}\label{EMo}
m\ddot{u}_n&=&
 2\alpha D[e^{-\alpha(u_{n+1}-u_{n}-r_m)}-e^{-2\alpha(u_{n+1}-u_{n}-r_m)}\nonumber \\
[4mm]
 &+&e^{-2\alpha(u_{n}-u_{n-1}-r_m)}-e^{-\alpha(u_{n}-u_{n-1}-r_m)}]\nonumber \\
[4mm]
 &-& H(u_n)2\pi A\sin(2\pi u_n).
\end{eqnarray}
Here $H(\eta)$ is the Heaviside step function.

\subsection{Initial conditions}
\label{IC}

The equations of motion Eq.~(\ref{EMo}) are integrated numerically for the initial coordinates
\begin{eqnarray}\label{InitCoord}
u_n=n-K-5, \quad && {\rm for} \quad n=1,...,K\,, \nonumber \\
u_n=n-K-1, \quad && {\rm for} \quad n=K+1,...,N\,,
\end{eqnarray}
and initial velocities
\begin{eqnarray}\label{InitVeloc}
\frac{du_n}{dt}=V_0, \quad && {\rm for} \quad n=1,...,K\,, \nonumber \\
\frac{du_n}{dt}=0, \quad && {\rm for} \quad n=K+1,...,N\,,
\end{eqnarray}
of the particles. With these initial conditions the initial distance between particles $K$ and $K+1$ is equal to 5, so that the molecule does not interact with the chain. The molecule moves toward the chain with the velocity $V_0$ and it starts to interact with the chain when they get closer.

The initial energy of the molecule is
\begin{eqnarray}\label{InitEner}
E_0=\frac{KmV_0^2}{2}\,.
\end{eqnarray}

\begin{figure}
\includegraphics[width=8.0cm]{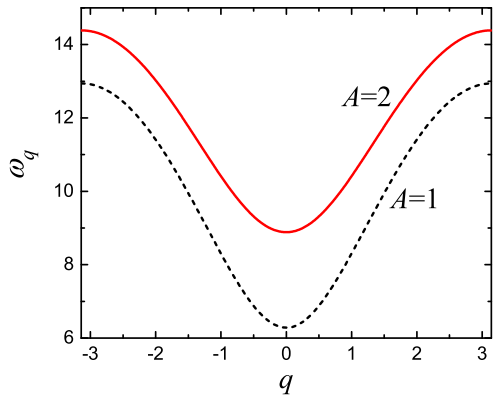}
\caption{Dispersion relation for the small-amplitude waves (phonons) supported by the considered chain of particles for two different values of the on-site potential depth, $A=1$ (dashed line) and $A=2$ (solid line).}
\label{fig2a}
\end{figure}

\subsection{Dispersion relation and phonon velocities}
\label{DPV}

In the case of small amplitude vibrations, the higher order nonlinear terms can be neglected, and Eq.~(\ref{EMo}) reduces to
\begin{equation} \label{EquationMotion}
m\ddot{u}_n=2\alpha^2D(u_{n-1}-2u_{n}+u_{n+1})-4\pi^2 Au_{n}.
\end{equation}
The solutions of the above equation are the linear combinations of
normal modes $u_n \sim \exp [i (q n -\omega_q t)]$ with
wave number $q$ and frequency $\omega_q$ obeying the dispersion relation
\begin{equation}\label{Dispersion}
\omega_q^2=\frac{4}{m}[\pi^2 A+\alpha^2 D(1-\cos q)].
\end{equation}

The dispersion relation~\eqref{Dispersion} is shown within the first Brillouin zone
in Fig.~\ref{fig2a} for $A=1$ (dashed line) and $A=2$ (solid line).
It suggests that the system supports the small-amplitude running waves (phonons)
with frequencies ranging from $\omega_{\min}=2\pi\sqrt{\frac{A}{m}}$ to
$\omega_{\max}=\frac{2}{\sqrt{m}}\sqrt{\pi^2A+2\alpha^2D}$.
Phonon's group velocity is defined by
\begin{eqnarray}\label{GroupVelocity}
v_g=\frac{\rm{d} \omega_{\mit{q}}}{\rm{d} \mit{q}}=\frac{\alpha^2D\sin q}{\sqrt{m}\sqrt{\pi^2A+\alpha^2D(1-\cos q)}}.
\end{eqnarray}
The group velocity vanishes for $q\rightarrow 0$ and $q\rightarrow \pm\pi$. For the considered model parameters, this function has a maximum value of $v_g^{\max}=3.3348$ ($v_g^{\max}=2.7504$) at $q=1.22$ ($q=1.33$) for $A=1$ ($A=2$).

\subsection{Static anti-kink (crowdion) in 1D model}
\label{Antikink}

Equilibrium anti-kink (crowdion) was obtained in 1D Frenkel-Kontorova model by setting initial atomic displacements with the help of the ansatz Eq.~(\ref{chainposition0}) with $a=1$, $\beta = 0.3$ and $x_0 = 41$ and subsequent relaxation. The resulting static anti-kink profiles are shown in Fig.~\ref{Kinks} by open squares and open circles for $A=1$ and $A=2$, respectively. Naturally, the anti-kink is narrower for deeper on-site potential, i.e., for $A=2$. With the chosen parameters, the anti-kinks in 1D model have width close to the width of crowdions in 3D Morse crystal, the latter ones are shown in Fig.~\ref{Kinks} by triangles and filled circles for two different relaxation procedures, as it was described in Sec.~\ref{Model3D}.

\section{Molecule bombardment}
\label{Numerics}

\begin{figure}
\includegraphics[width=8.0cm]{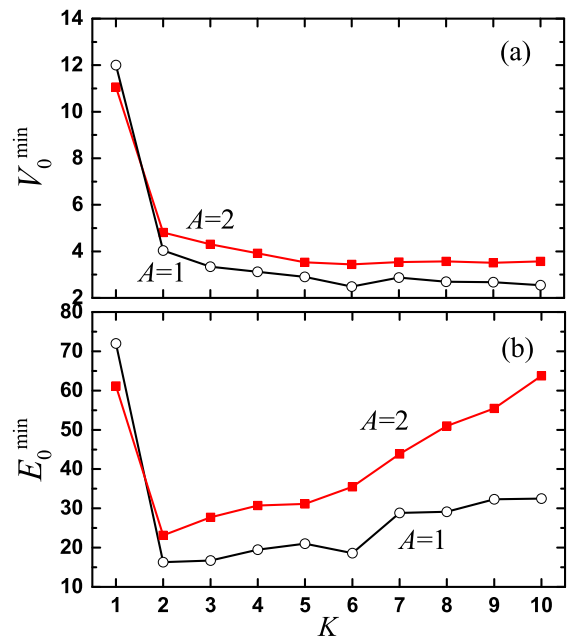}
\caption{Minimal velocity of the molecule of $K$ atoms required to initiate a kink by hitting the chain at the end. Results for the sinusoidal potential amplitude $A=1$ ($A=2$) are shown in black (red).}
\label{fig3a}
\end{figure}


\begin{figure}
\includegraphics[width=8.0cm]{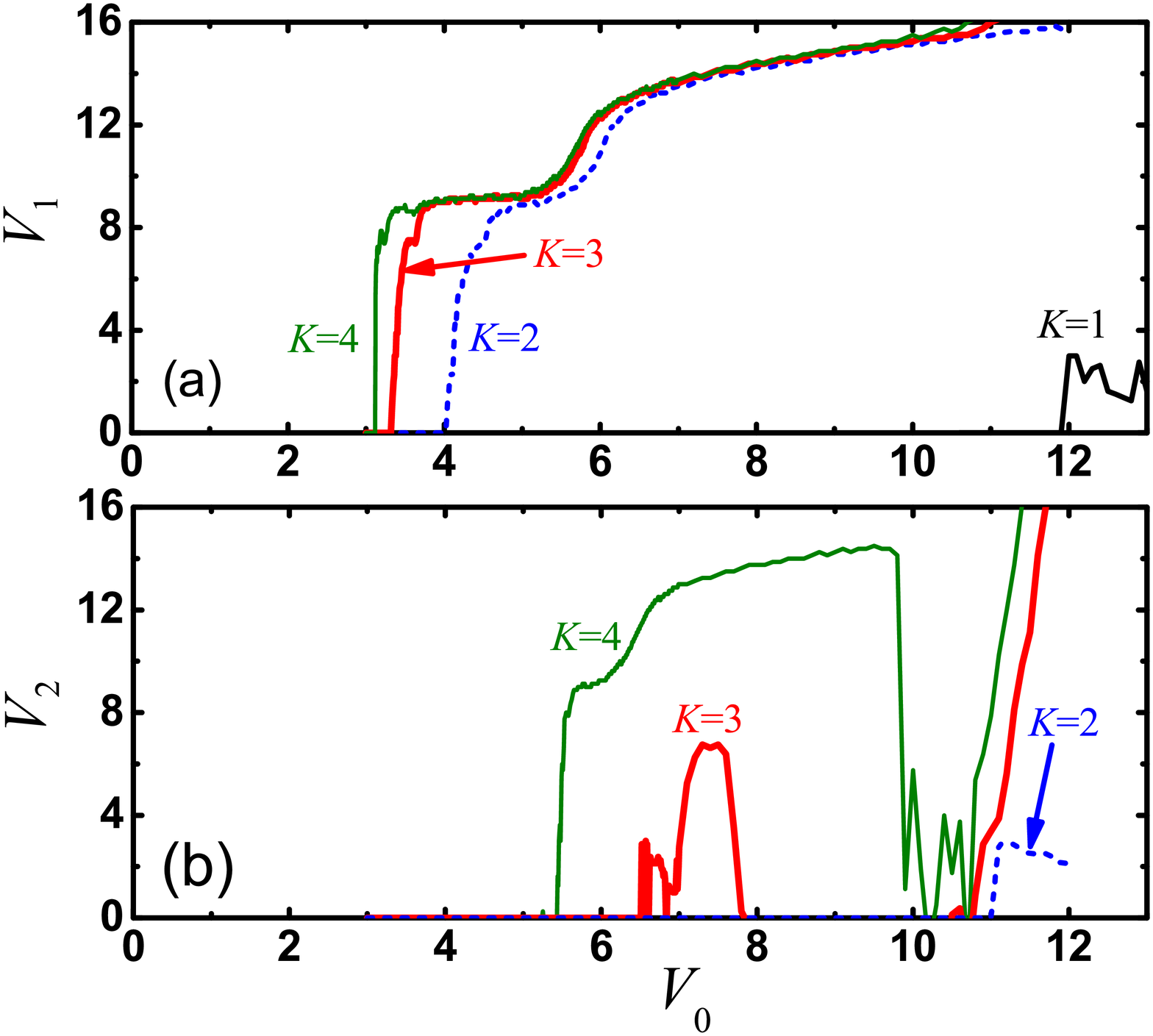}
\caption{Velocity of the fastest (a) and second fastest (b) crowdions created in the chain by molecule bombardment as a functions of the molecule initial velocity. The case of single atom ($K=1$) is shown by black line. Results for the molecules with $K=2$, 3, and 4 atoms are shown by blue dashed, thick red, and thin green lines, respectively.}
\label{fig4}
\end{figure}
\begin{figure}
\includegraphics[width=8.0cm]{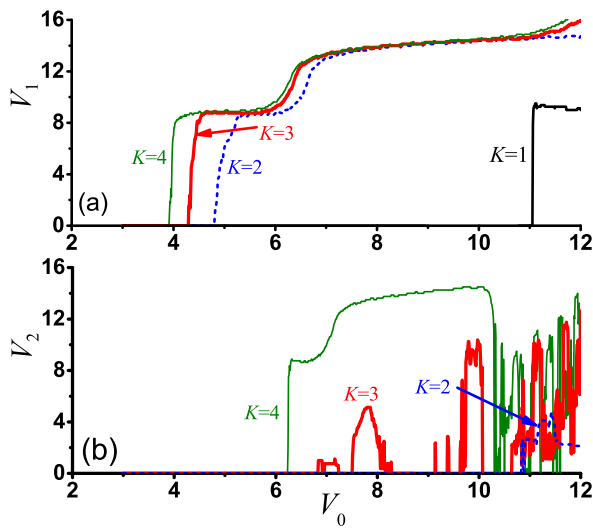}
\caption{Same as in Fig.~\ref{fig4}, but for the on-site potential depth $A=2$. }
\label{fig5}
\end{figure}

Let us discuss the results of numerical simulation of molecule bombardment.

Firstly we find the minimal velocity $V_0^{\min}$ of the molecule of $K$ atoms needed to initiate a crowdion. With the use of Eq.~(\ref{InitEner}), we calculate the corresponding minimal energy of the molecules $E_0^{\min}$ required to produce a crowdion. The results are presented in Fig.~\ref{fig3a} for the on-site potential depth $A=1$ (open circles) and $A=2$ (squares). It is clear that single atom ($K=1$) needs higher initial velocity to launch mass transport along the chain as compared to the molecules ($K>1$). Minimal initial energy for $K=1$ is higher than that for $2\le K \le 9$ in the case of $A=2$, and even for longer molecules for $A=1$.

This effect can be understood taking into account the fact that static (or slowly moving) crowdion has a width of half a dozen of atoms, see Fig.~\ref{Kinks}. Single atom cannot produce a wide, slowly moving crowdion but it can only produce relatively sharp fast-moving crowdions, which requires high energy. Already molecule with $K=2$ atoms has a non-zero size and it is much more efficient in initiation of crowdion. Indeed, for $A=1$ ($A=2$) the molecule of two atoms needs 4.4 (2.7) times smaller energy than single atom to initiate a crowdion. The reduction of energy required to create a crowdion is more pronounced for shallower on-site potential, i.e., for $A=1$. This is because the crowdion is wider for $A=1$ and it is more difficult for a single atom to produce it.

Note a local minimum of $E_0^{\min}$ at $K=6$ for the case of $A=1$ in Fig.~\ref{fig3a}(b). This is because the molecule of this size is compatible with the crowdion width. For the case of $A=2$, a similar tendency of reduction of $E_0^{\min}$ is observed for $K=5$ and 6, in line with the fact, that the crowdion width in this case is somewhat smaller than for $A=1$.

We have also calculated speed of crowdions moving along the chain as a function of the initial molecule velocity $V_0$ for molecules with $K=1$, 2, 3, and 4. For $A=1$ ($A=2$) the results are shown in Fig.~\ref{fig4} (Fig.~\ref{fig5}). For sufficiently large $V_0$, more than one crowdion can be initiated by the molecule impact, and we plot velocities of the first and second fastest crowdions in (a) and (b), respectively. Clearly, molecules with $K>1$ need considerably smaller minimal initial velocity to produce crowdions as compared to single atom. Molecules initiate crowdions propagating at a higher speed. Note that crowdions have preferable propagation velocities and this issue will be addressed in Sec.~\ref{KinkSolution}. Within the studied range of initial velocities $V_0$, molecules are able to produce more than one crowdion, in contrast to single atom.

Overall, we conclude that bombardment with molecules is much more efficient in initiation of mass transport along the chain than bombardment with single atoms.

%

\section{Crowdions in 1D chain}
\label{KinkSolution}

In order to better understand the results presented in Sec.~\ref{Numerics}, we analyse here properties of crowdions in the considered Frenkel-Kontorova chain.
Firstly, we derive the moving crowdion solution under the assumption of harmonic interatomic coupling. Unfortunately, this solution is valid only for very wide crowdions, but not for the crowdions spanning over a half a dozen of atoms considered here. That is why we then study crowdions numerically for the chain with Morse interatomic interactions.

\subsection{Analytical treatment}
\label{KinkSolutionAnalit}
In the long-wave approximation, $|u_{n+1}-u_n|\ll 1$ for all $n$, Eq.~(\ref{EMo}) can be simplified by linearizing the interatomic forces.
This results in the Frenkel-Kontorova model with harmonic inter-particle interactions
\begin{equation} \label{KinkMotion}
m\ddot{u}_n=2\alpha^2D(u_{n-1}-2u_{n}+u_{n+1})-2\pi A\sin(2\pi u_n).
\end{equation}
Introducing the new variable
\begin{equation} \label{NewVar}
w_n=2\pi u_n,
\end{equation}
we rewrite Eq.~(\ref{KinkMotion}) in the form
\begin{equation} \label{KinkMotionNew}
\ddot{w}_n=\frac{1}{h^2}(w_{n-1}-2w_{n}+w_{n+1})-g^2\sin(w_n),
\end{equation}
where
\begin{equation} \label{Params}
h^2=\frac{m}{2\alpha^2D}, \quad g^2=\frac{4\pi^2 A}{m}.
\end{equation}
In the continuum limit, $h \rightarrow 0$, Eq.~(\ref{KinkMotionNew}) reduces to the sine-Gordon equation
\begin{equation}\label{SGE}
w_{tt}-w_{xx}+ g^2\sin w=0,
\end{equation}
which has the well-know moving kink solution
\begin{equation}\label{Kink}
w(x,t)=2\pi\pm 4\arctan\Big[\exp\Big(g\frac{x-x_0-vt}{\sqrt{1-v^2}}\Big)\Big],
\end{equation}
where $v$ defines the kink velocity and $x_0$ represents its initial position.
For the upper (lower) sign we actually have kink (crowdion) solution.

Returning to the original variable $u_n$ and taking into account $x=nh$, we write the approximate kink solution to Eq.~(\ref{KinkMotion}) in the following form
\begin{equation}\label{KinkOrigin}
u_n(t)=1\pm \frac{2}{\pi}\arctan\Big[\exp\Big(g\frac{h(n-x_0)-vt}{\sqrt{1-v^2}}\Big)\Big].
\end{equation}
From this solution, actual kink velocity is equal to $V_{\rm kink}=v/h$. Since $|v|<1$, the kink velocity is within the range
\begin{equation}\label{MaxVel}
|V_{\rm kink}|<\frac{1}{h}.
\end{equation}
From Eq.~(\ref{Params}), we find that for parameters used in our study $|V_{\rm kink}|<5.66$.

According to the solution Eq.~(\ref{KinkOrigin}), maximal absolute value of the kink's slope is
\begin{equation}\label{KinkSlope}
\beta=\frac{gh}{\pi\sqrt{1-v^2}}=\sqrt{\frac{2A}{\alpha^2D(1-v^2)}}.
\end{equation}
The maximal kink slope diverges as $|v|\rightarrow 1$, which means the kink width vanishes in this limit.

It is worth noting that kinks and crowdions of the Frenkel-Kontorova model Eq.~(\ref{KinkMotion}) have the same maximal slope, but this is not the case for the kinks and crowdions of the model Eq.~(\ref{EMo}) because stiffness of the Morse potential increases (decreases) under compression (tension) of the chain, while stiffness of the linear bonds is strain independent.

\subsection{Numerical results}
\label{KinkSolutionNumerical}

\subsubsection{Static kinks and crowdions}
\label{StaticKink}

\begin{figure}
\includegraphics[width=8.0cm]{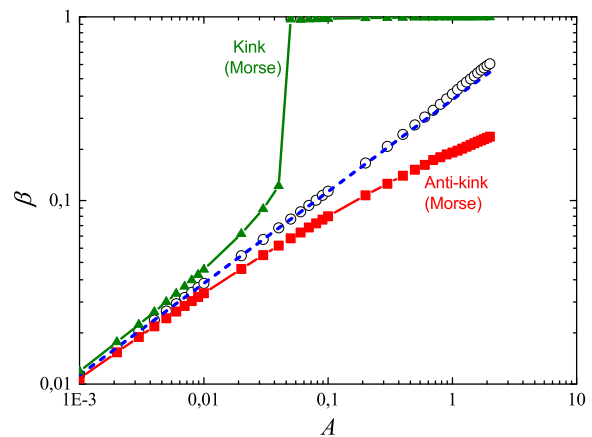}
\caption{Maximal absolute value of the slope of the static crowdion $(v=0)$ as a function
of the on-site potential amplitude. Straight dashed line gives the analytical solution Eq.~(\ref{KinkSlope}).
Open circles are for the numerical solution of Eq.~(\ref{KinkMotion}) both for kinks and crowdions.
Red squares (green triangles) are for the crowdion (kink) obtained numerically by solving Eq.~(\ref{EMo}).}
\label{fig3}
\end{figure}

Let us assess the accuracy of the kink solution Eq.~(\ref{KinkOrigin}) by calculating numerically the maximal slope of the static crowdion, $\beta$, as a function of the on-site potential depth, $A$. To do so, the chain of $N=1000$ particles is considered with the kink placed in the middle using Eq.~(\ref{KinkOrigin}) with $x_0=500.5$ and $v=0$. With this choice of $x_0$ the inter-site kink is obtained which has lower energy as compared to the on-site kink. In order to find equilibrium positions of the atoms, viscosity was introduced in the system by adding the term $\gamma\dot{u}_n$ to the left-hand side of Eq.~(\ref{EMo}) and Eq~(\ref{KinkMotion}) with the viscosity coefficient $\gamma=0.1$. Simulation run is carried out until maximal force acting on atoms becomes less than $10^{-12}$. Then the maximal crowdion slope is calculated as the slope of the line connecting two central atoms of the kink (crowdion).

The results for the relaxed kinks are presented in Fig.~\ref{fig3} using log-log scale. The straight dashed line shows the analytical solution Eq.~(\ref{KinkSlope}), while open circles are for the numerical solution of Eq.~(\ref{KinkMotion}). In the case of harmonic interparticle bonds, as it was mentioned earlier, kink and crowdion have same $\beta$. This is not true for Eq.~(\ref{EMo}) with Morse interatomic interactions. For this case the numerical results for crowdion (kink) are shown by red squares (green triangles). Note a sharp increase of $\beta$ when $A$ exceeds 0.04 for the Morse kink. For larger $A$ the kink transforms into a vacancy since one of the Morse bonds breaks.

It can be seen from Fig.~\ref{fig3} that the analytical solution Eq.~(\ref{KinkOrigin}) gives
a very good estimation of the kink maximal slope in the case of harmonic interatomic interactions,
Eq.~(\ref{KinkMotion}), within the entire studied range of $10^{-3}\le A\le 2$. However,
for the original model with Morse interactions, Eq.~(\ref{EMo}), the analytical solution
can be used only for $A<10^{-2}$, where the relative error in estimation of $\beta$ is less than 10\%.
The reason of such a poor accuracy is strong nonlinearity of the Morse potential.
As it was shown in Sec.~\ref{Model3D}, for 3D Morse crystal $A=1.22$, and in this
case the nonlinearity of the interparticle bonds cannot be neglected.

\begin{figure}[h!]
  \centering
  \includegraphics[width=0.5\textwidth]{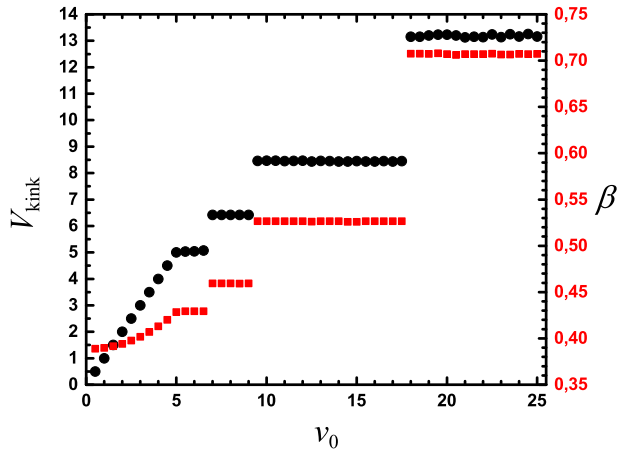}\label{Vel1}
    \caption{Velocity $V_{\rm kink}$ (circles) and parameter $\beta$ (squares) of the crowdion in the regime of steady motion as a functions of the initial velocity $v_0$ in the ansatz Eq.~(\ref{chainposition}). On-site potential depth is $A=1$.}
  \label{Steady1}
\end{figure}

 \begin{figure}[h!]
  \centering
  \includegraphics[width=0.5\textwidth]{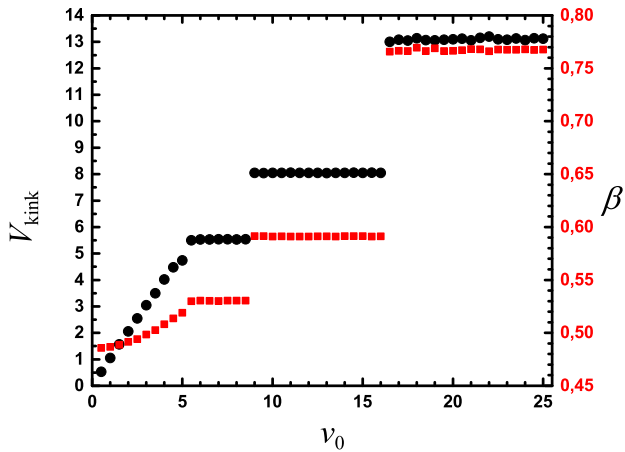}\label{Vel2}
    \caption{Same as in Fig.~\ref{Steady1} but for $A=2$.}
  \label{Steady2}
\end{figure}

\begin{figure}[h!]
  \centering
  \includegraphics[width=0.5\textwidth]{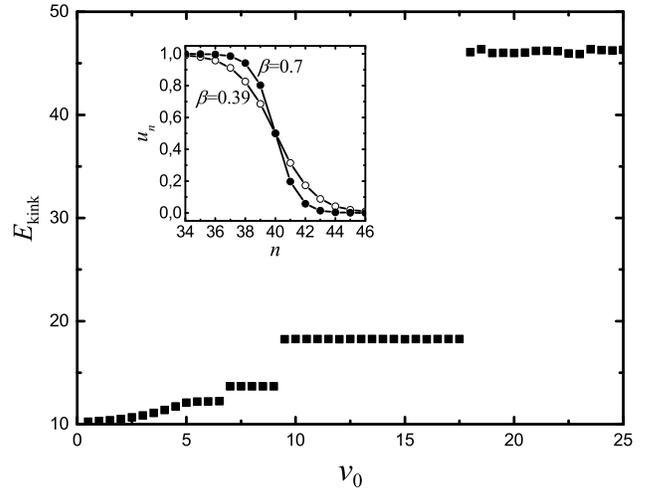}\label{Vel1}
    \caption{Energy of crowdion in the regime of steady motion as a functions of the initial velocity $v_0$ in the ansatz Eq.~(\ref{chainposition}). On-site potential depth is $A=1$. Inset shows kink profiles for $\beta=0.39$ (open dots) and $\beta=0.7$ (filled dots).}
  \label{EkinkA1}
\end{figure}

 \begin{figure}[h!]
  \centering
  \includegraphics[width=0.5\textwidth]{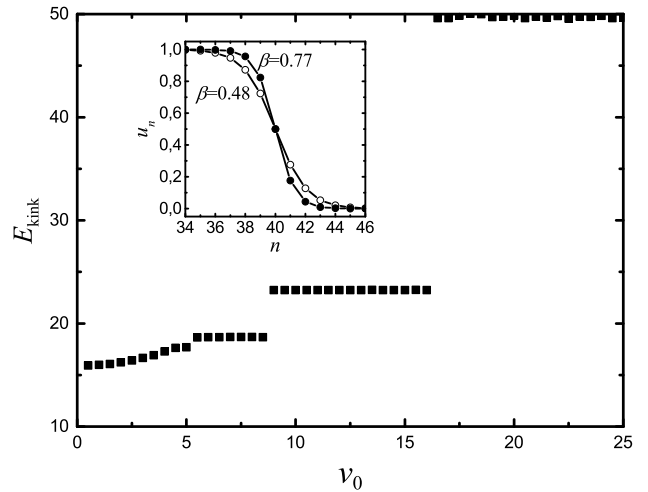}\label{Vel2}
    \caption{Same as in Fig.~\ref{EkinkA1} but for $A=2$. Inset shows kink profiles for $\beta=0.48$ (open dots) and $\beta=0.77$ (filled dots).}
  \label{EkinkA2}
\end{figure}

\subsubsection{Moving crowdions}
\label{MovingKink}

For initiation of moving crowdions in 1D Frenkel-Kontorova model Eq.~(\ref{EMo}), the following ansatz is adopted
\begin{eqnarray}\label{chainposition}
u_{n}(t)&=&\frac{1}{2}-\frac{1}{2}\tanh[\beta_0(n-x_0-v_0t)],
\end{eqnarray}
where $\beta_0$, $v_0$, and $x_0$ are the initial crowdion inverse width, velocity, and position, respectively.
In the chain of 2000 particles, at $t=0$, crowdion moving with a positive velocity is excited at the site $x_0=100$. The range of crowdion initial velocities $0 < v_0 \le 25$ is studied. For a chosen value of $v_0$, parameter $\beta_0$ is found by the trial and error method aiming to achieve a minimal radiation from the moving crowdion. At the end of numerical run at $t=50$, the crowdions achieve a state of steady motion. The steady crowdion velocity $V_{\rm kink}$ and inverse width $\beta$ are measured. The latter parameter is found by the least squares fitting of the kink profile to the expression Eq.~(\ref{chainposition}).

The results for steadily moving crowdions are presented in Fig.~\ref{Steady1} for the on-site potential depth $A=1$ and in Fig.~\ref{Steady2} for $A=2$. Shown are $V_{\rm kink}$ (circles, left scale) and $\beta$ (squares, right scale) as a functions of the initial velocity $v_0$ in the ansatz Eq.~(\ref{chainposition}).

Interestingly, two different regimes are observed for slow and fast crowdions. When $v_0<5$, we have $V_{\rm kink}=v_0$, but faster crowdions can have only selected velocities. For $A=1$, within the studied range of $v_0$, selected velocities are 5.06, 6.42, 8.45, and 13.2. For $A=2$, they are 5.53, 8.05, and 13.1. This explains the plateaus observed in Figs.~\ref{fig4} and \ref{fig5} at the velocities around 8 and 13. Note that the value of velocity separating two different regimes is close to the estimation of maximal kink velocity that follows from the approximate solution reported in Sec.~\ref{KinkSolutionAnalit}, see the text below Eq.~(\ref{MaxVel}). In fact, kinks propagating in nonlinear chains with selected velocities have been reported in a number of studies, e.g., in \cite{Select1,Select2,Select3,Select4,Select5,Select6,Select7}.

Similarly, crowdion inverse width $\beta$ monotonically increases for $0< v_0<5$, but it has discrete values for faster crowdions.

It is also instructive to analyse total (kinetic plus potential) energies of antikinks, $E_{\rm kink}$, in the regime of steady motion as the functions of $v_0$. For $A=1$ ($A=2$), the result is shown in Fig.~\ref{EkinkA1} (Fig.~\ref{EkinkA2}). Note that slow crowdions ($v_0<5$) have relatively small energy, in the range $10.3<E_{\rm kink}<12$ for $A=1$ and $15.9<E_{\rm kink}<17.7$ for $A=2$. Faster crowdions have considerably higher energy. Insets in Figs.~\ref{EkinkA1} and \ref{EkinkA2} show the kink profiles for the smallest and largest value of $\beta$ observed in the studied range of parameter $0<v_0\le 25$.

Notably, phonon velocities do not play an essential role in crowdion propagation in 1D chain. Maximal phonon velocities were estimated below Eq.~(\ref{GroupVelocity}) to be about 3. In Figs.~\ref{Steady1} to \ref{EkinkA2}, we do not see any peculiarities around this velocity. In 2D and 3D crystals, sound velocity does play an important role in crowdion motion \cite{Nc1,Nc2} because the close-packed atomic row in which crowdion propagates interacts with the surrounding atoms.

Presented results contribute to understanding why bombardment with molecules ($K>1$) produces crowdions in the chain more efficiently than single atoms ($K=1$). As it was pointed out in Sec.~\ref{Numerics}, slow crowdions are rather wide and they cannot be excited by single atoms. Fast crowdions are narrower and they can be excited by single atoms, but it requires sufficiently large energy. On the other hand, low-energy slow kinks can be excited by molecules since they have nonzero size.

\section{Conclusions}
\label{Conclusions}

We have simulated mass transfer in Frenkel-Kontorova chain by anti-kinks (crowdions) initiated by single atom or molecule bombardment. Parameters of the Frenkel-Kontorova chain with Morse interatomic interactions were chosen to mimic crowdions in 3D fcc Morse crystal. Our main results can be summarized as follows.

Static or slowly moving crowdions have width of about half a dozen of atoms. Such wide crowdions cannot be initiated by an impact of single atom due to a mismatch of their sizes. Atoms can initiate only fast crowdions because they are narrow, but this requires relatively large energy. Molecules can initiate wide and slow crowdions with small energy, since they have size compatible with the crowdion width. That is why molecules require three to four times smaller energy to initiate a crowdion propagating along the chain as compared to minimal energy of single atom needed for this.

Our findings are of importance for experimental techniques where atom or molecule bombardment is used to modify properties of crystal surface.

As a continuation of this study is would be interesting to estimate the efficiency of molecule bombardment in mass transfer initiation for real crystals in 3D setting.

\begin{acknowledgments}
For E.A.K. this work was supported by the Russian
Science Foundation, grant No. 1612-10175 (discussion of the numerical results and writing the paper). The research of E.G.S. is carried out at Tomsk Polytechnic University within the framework of Tomsk Polytechnic University Competitiveness Enhancement Program grant (discussion of the numerical results). S.V.D. thanks the Russian Foundation for Basic Research, grant No. 17-02-00984-a (statement of the problem and discussion of the numerical results). 
\end{acknowledgments}


\end{document}